\documentclass[12pt]{article}
\usepackage{multido}
\usepackage{mathrsfs}
\usepackage{amsmath}
\usepackage{amssymb}
\usepackage{amsthm}
\usepackage{dsfont}
\usepackage{amsfonts}
\usepackage{fullpage}		
\usepackage{fancyhdr}
\usepackage{graphicx}
\usepackage{units}
\usepackage{setspace}
\usepackage[usenames,dvipsnames]{color}
\usepackage{appendix}
\usepackage{slashed}
\def\ten{\mathcal T}
\def\mass{\mathcal M}
\def\vbar{\overline{ v}}

\def\bnabla{\bar\nabla}

\begin{document}
\begin{titlepage}
\vfill
\begin{flushright}
\end{flushright}

\vfill
\begin{center}
\baselineskip=16pt
{\Large\bf Gravitational Tension and Thermodynamics}
\vskip 0.3in
{\Large\bf  of  Planar AdS Spacetimes}
\vskip 1.0cm
{\large {\sl }}
\vskip 10.mm
{\bf Basem Mahmoud El-Menoufi\footnote{bmahmoud@physics.umass.edu}, Benjamin Ett\footnote{bett@physics.umass.edu}, David Kastor\footnote{kastor@physics.umass.edu},
 Jennie Traschen\footnote{traschen@physics.umass.edu}} \\
\vskip 1cm
{
Department of Physics\\
University of Massachusetts\\
Amherst, MA 01003
     	     }
\vspace{6pt}
\end{center}
\vskip 0.5in
\par
\begin{center}
{\bf Abstract}
 \end{center}
\begin{quote}

We derive new thermodynamic relations for  asymptotically planar AdS black hole and soliton
solutions. 
In addition to the ADM mass, these
 spacetimes are characterized by gravitational tensions in each of the planar spatial directions. We show that 
with planar AdS asymptotics, the sum of the ADM mass and tensions necessarily vanishes, 
as one would expect from the AdS /CFT correspondence.
Each Killing vector of such a spacetime leads
to a Smarr formula relating the ADM mass and tensions, the black hole
horizon and soliton bubble areas, and a set of thermodynamic volumes that arise due to the non-vanishing cosmological constant. These Smarr relations display
 an interesting symmetry between black holes and
bubbles, being invariant under the simultaneous interchange of the mass and black hole horizon area with the
tension and soliton bubble area.  This property may indicate a symmetry between the confining and deconfined phases of the dual gauge theory.

\vfill
\vskip 2.mm
\end{quote}
\end{titlepage}
\section{Introduction}

This paper will explore certain aspects of the thermodynamics of asymptotically planar AdS spacetimes, such as the planar black hole
 \cite{Lemos:1994xp,Cai:1996eg,Horowitz:1998ha} and soliton \cite{Horowitz:1998ha} solutions studied in these references.
Through the gauge/gravity correspondence, the thermodynamics of such spacetimes are known to have important physical consequences.
For example, it was argued in \cite{Horowitz:1998ha}  following \cite{Witten:1998zw,Witten:1998qj}
that the planar AdS black hole is
 dual to the high temperature, deconfined phase of the associated
 strongly coupled gauge theory  on the AdS boundary, with the AdS soliton corresponding to the
 low temperature, confining phase of the theory. 
 Further support for this interpretation comes from
  the values of the Euclidean actions, or free energies,
  of the  black hole \cite{Peca:1998dv,Surya:2001vj}  and the soliton 
 \cite{Surya:2001vj} which demonstrate
that the system  indeed undergoes
a phase transition as in \cite{Hawking:1982dh}. Further studies of {\it e.g.} the approach to equilibrium in this system have appeared in 
\cite{de Mello Koch:1998qs,Horowitz:1999jd,Hubeny:2009rc,Haehl:2012tw}.

The gravitational charges of planar AdS spacetimes display interesting features, with the AdS soliton having a negative, but nonetheless physical, mass \cite{Horowitz:1998ha}. Our exploration begins with the observation that 
the mass does not fully characterize the asymptotic behavior of such spacetimes.
Due to the planar spatial directions at infinity one expects that, as in the case for Kaluza-Klein black holes 
\cite{Harmark:2003dg,Elvang:2004iz,Kastor:2006ti,Kastor:2008wd},
gravitational tensions should also make important contributions to basic thermodynamic relations. 
Indeed, the tracelessness of the boundary CFT stress tensor implies that tension or pressure\footnote{Tension is defined as the negative of pressure.} must play an important role.
Consistent with this, we will see that the AdS soliton, with its negative mass, has a large positive tension, with this situation
being reversed for the black hole. 

With nonzero $\Lambda$ another important  ingredient enters the thermodynamic relations as well.  It has been shown that to derive a Smarr formula for
(A)dS black holes, one must include a new potential $V$, which is thermodynamically conjugate to the cosmological constant in the sense that it gives the change in mass when 
$\Lambda$ 
is varied \cite{Kastor:2009wy}.  This new potential $V$ has the dimensions of volume, and following \cite{Cvetic:2010jb} we will  refer to it as the thermodynamic volume.  Regarding the cosmological constant as a pressure, the Smarr formula for (A)dS black holes then contains a pressure-volume type contribution 
proportional to $\Lambda\, V$.
Computations of $V$ and studies of associated thermodynamic phenomena 
have been carried out in \cite{Kastor:2009wy,Cvetic:2010jb,Kubiznak:2012wp,Dolan:2012jh,Dolan:2011xt,Dolan:2010zz} for AdS black holes with spherical topology, and in \cite{Dolan:2013ft} for the dS case.

Smarr formulas for gravitational systems are analogous  to the Gibbs-Duhem relations of 
classical thermodynamics. In the gravitational case, a Smarr formula is
 an exact relation that holds for stationary black holes, and more generally for any spacetime having a Killing symmetry, that  relates the
ADM  gravitational charges \cite{Arnowitt:1962hi} defined in an asymptotic region to geometric properties of the interior of the spacetime, such as the area and surface gravity of a black hole horizon. 
Originally derived for four dimensional, stationary,
asymptotically flat black holes ({\it i.e.} black holes of  astrophysical interest) in \cite{Smarr:1972kt},
extensions of these relations have been made
to higher spatial dimensions, different horizon topologies, higher derivative theories of gravity, and
to spacetimes with different asymptotics. 

We will
derive Smarr relations for asymptotically  (locally) AdS spacetimes with planar spatial cross sections, {\it i.e.} such that 
  the boundary  at infinity is a plane or torus cross time.  The  
 ADM charges for these spacetimes are the mass and spatial tensions together with the linear and angular momenta. We will focus on
 static spacetimes, with vanishing linear and angular momenta,  containing
 black holes and/or bubbles.
  This problem appears at first sight
to be a simple generalization of two previously considered cases. Black holes and bubbles
 with compact spatial directions (but $\Lambda=0$) were considered in  \cite{Elvang:2004iz,Elvang:2002br,Kastor:2008wd,Harmark:2003dg,Kastor:2006ti},  while spherical AdS black holes ({\it i.e.} with $\Lambda\neq 0$ but no compact spatial directions) were considered in \cite{Bazanski:1990qd,Kastor:2008xb}. The compact spatial directions introduce new terms in the Smarr
 relations of the form $ L\,\ten$, 
 with $L$ being the length of a given compact spatial direction and $\ten$ being the gravitational tension
in that direction, while as noted above taking $\Lambda\neq 0$ introduces  a $ \Lambda\, V$ term. We will see, however, that combining
 compact spatial dimensions
with $\Lambda\neq 0$ is in fact not just a simple addition problem, but rather involves interesting new features.

The most prominent new feature that arises is a tracelessness condition
\begin{equation}\label{zero}
\mass +\sum _i  L_i\ten _i =0
\end{equation}
satisfied by the ADM mass and tensions, where the index $i$ runs over the planar spatial directions.
While a condition of this sort would naturally be expected from the AdS/CFT standpoint, it is interesting to see how it arises
on the gravity side in terms of standard ADM conserved charges \cite{Arnowitt:1962hi}. The condition (\ref{zero}) also helps make sense of  unusual signs 
that arise for the mass and tensions with $\Lambda<0$.   In the asymptotically flat case, the ADM mass is known to be positive \cite{Schon:1981vd,Witten:1981mf},
and the gravitational tensions have also been shown to be positive \cite{Traschen:2003jm}
 for static spacetimes.
However, as we will see, in AdS there are examples with positive mass  but negative tensions, and similarly solutions with positive tensions but negative mass.  Equation (\ref{zero}) makes it clear this is to be expected in AdS.  All the gravitational charges cannot be positive.
The constraint (\ref{zero}) also
implies that the ADM charges are not all independent, which will have practical implications in our construction. In particular,  
the inversion of Komar integral identities necessary to obtain the Smarr results will involve technical complications, dealt with in section (\ref{technical}) below, that were not present in the $\Lambda=0$ case.
 
Our strategy in the paper is as follows.  We study solutions to the Einstein equation with negative cosmological
 constant
\begin{align}\label{einstein}
R_{ab}=\frac{2\Lambda}{D-2}g_{ab}
\end{align}
where $\Lambda = -(D-1)(D-2)/2l^2$.
We will assume that the spacetime is static and has $p$ spatial
translation Killing vectors with $0\leq p\leq D-2$. The planar
black hole  and AdS soliton
solutions studied in \cite{Horowitz:1998ha} both have $D-2$ spatial translation symmetries. We will also
assume that there are no shear terms at infinity,
so that asymptotically the metric has the form
\begin{align}\label{adsfalloff}
ds^2 \simeq  {r^2 \over l^2 } \left(\eta_{\alpha\beta} dx^\alpha dx^\beta 
+\sum_{\alpha=0}^{D-2} {c_\alpha \over r^{D-1}}  (dx^\alpha)^2 \right) 
+
{l^2 \over r^2}\left(1 + {c_r \over r^{D-1}}\right)  dr^2 
\end{align}
where the $c_\alpha$'s and $c_r$ are constant fall-off coefficients, and we take the planar spatial directions $x^i$ with $i=1,\dots, D-2$ 
to be periodically identified according to
$x^i\equiv x^i+L_i$.
We assume that $p+1$ of the vector fields ${\xi} _{(\beta)}=\partial/\partial x^\beta$, including necessarily ${\xi} _{(0)}$, satisfy Killing's equations throughout the spacetime.
Each of these Killing vectors leads to a Komar integral identity, which provide relations between the the fall-off coefficients $c_\alpha$,
and geometric properties of black hole horizons and/or soliton bubbles in the interior, plus terms of the form  $\Lambda V_{(\beta)}$,  where
$V_{(\beta)}$ is the thermodynamic volume associated with the Killing vector ${\xi} _{(\beta)}$.
 After taking into account the trace constraint (\ref{zero}) and
a related gauge fixing, the fall-off coefficients $c_\alpha$ may be solved for in terms of the ADM charges giving the Smarr relations. 

There is a geometric distinction between
spacelike Killing vectors that act freely and those that have one or more fixed points, referred  to as  `bubbles'  (see {\it e.g.}  \cite{Elvang:2004iz,Elvang:2002br,Harmark:2003dg} for related discussions of bubbles with $\Lambda=0$).
Gravitational tensions then divide into three categories; tensions associated with freely acting symmetry directions, which will be denoted by $\ten_{F}$; 
tensions associated with non-freely acting symmetry directions, which will be denoted\footnote{Further notation will be added as needed below to distinguish between different freely, or non-freely, acting directions.} by $\ten _{B}$; and those which
are not along a symmetry direction.   
The asymptotic lengths of these dimensions will similarly be denoted by $L_F$ and $L_B$.
The Smarr relations are simplest when all the Killing vectors commute, in which case we will 
show that all the thermodynamic volumes are equal, $V_{(\beta )}= V$, and 
the Smarr relations reduce to
\begin{eqnarray}\label{introsmarr}
\mass & =&    {\kappa_HA_H \over 8\pi} + \frac{   \Lambda V}{ 8\pi (D-2) } \\ 
   L_{B} \ten _{B}    &= & 
  {\kappa_B L_B A_{B} \over 8\pi} + \frac{   \Lambda V}{ 8\pi (D-2) } 
\\ 
L_F \ten_F  &=& \frac{   \Lambda V }{ 8\pi (D-2) }
\end{eqnarray}
where the surface gravity for a smooth bubble is given by $\kappa_B L_B=2\pi$ (see \cite{Kastor:2008wd} for further discussion). The formula for $\ten_F $
is a special case of  $\ten_B$ with the bubble area equal to zero.
In this case the formula says that the globally defined 
 gravitational tension times the length, $L_F\ten_F$, is proportional to minus the intrinsic pressure  times the thermodynamic volume $V$, where the intrinsic pressure is due to
 the cosmological constant with $P\propto -\Lambda$.
The Smarr formulas display an interesting symmetry between black holes and bubbles, since
they are invariant under interchanging the contributions from 
  the mass and horizon area with the contributions of the bubble tension and bubble area. 
These  equations also raise the intriguing  possibility of
a critical solution  that is a mixture of bubbles and black holes,
representing a transition between the high and low temperature phases of the field theory, which would have
 $\mass = \ten_{B} =0$, $\kappa_H A_H / 8\pi =
\kappa_BA_{B} /8\pi =-  \Lambda V / 8\pi (D-2)  $, and positive tensions
which are not associated with a symmetry to satisfy the trace constraint (\ref{trace}).

This paper is organized as follows. In Section (\ref{ADM}) we use the ADM formalism to compute the mass and gravitational tensions in terms of the assumed asymptotic form of the metric  (\ref{adsfalloff}). In Section (\ref{trace}) we show how to take into account the trace condition 
(\ref{zero}) amongst the gravitational charges 
and deal with a related gauge fixing issue.   We also compare the ADM charges to the results of boundary stress tensor 
computations. The Smarr relations are derived in Section (\ref{smarrsection})
and the results are collected in Section (\ref{results}) along with 
 comments on open questions. The Appendix contains a derivation.
We use the notational conventions that Greek indices $\alpha , \beta= 0, \dots , D-2$ label directions tangent
  to the planes of constant radial coordinate $r$, and Latin indices $i,j=1,\dots , D-2$ label the spatial directions within these planes.

\section{Mass and tensions}\label{ADM}

In this section, we find expressions for the ADM mass and tensions in terms of the
fall-off coefficients $c_\alpha$ and $c_r$ that characterize the asymptotic form of the metric in (\ref{adsfalloff}). We begin by
reviewing the ingredients of this calculations in 
asymptotically AdS spacetimes following
 \cite{Abbott:1981ff,Kastor:2009wy}. 
We then evaluate the mass and tensions for the planar AdS black hole metric and the AdS soliton
\cite{Horowitz:1998ha}, which will serve as important examples.

\subsection{ADM charges}
An ADM gravitational charge $Q(\xi)$ is defined with respect to an asymptotic symmetry $\xi^a$
of a spacetime \cite{Abbott:1981ff}.  The first step in the computation is then to specify
the boundary conditions on the metric and hence the asymptotic symmetries.   As noted, we consider spacetimes that are
asymptotically  AdS in Poincare coordinates, with the rate of fall-off specified by equation (\ref{adsfalloff}).
 These conditions are analogous to standard asymptotically flat boundary conditions
 in the sense that they
give finite mass and tensions, with values that are independent of the choice of  slice 
at infinity used to compute the charge. Other choices of  asymptotics for AdS may also be of interest, such as including logarithmic  terms in the far field, or allowing the
fall off coefficients  to depend on the planar coordinates, but we defer
consideration of these to future work. 

The construction of ADM charges can be nicely expressed in the Hamiltonian formalism \cite{Regge:1974zd}. 
We will follow the particular treatment given in  \cite{Traschen:2001pb}, which we briefy recount here.  Let the 
 spacetime  be foliated by a family of hypersurfaces $\cal{S}$ with unit normal $n^a$, and 
 decompose the metric as 
\begin{equation}\label{foliation}
g_{ab}=(n \cdot n)n_an_b+s_{ab} \   , \quad n \cdot n=\pm 1 \   , \quad n^as_{ab}=0 ,
\end{equation}
where $s_{ab}$ is the induced metric on $\cal{S}$ with  conjugate
 momentum $\pi ^{ab}$ .
  Let ${\bar s}_{ab}$ and $ {\bar \pi}^{ab}$ denote the metric and momentum in AdS, and let
 $h_{ab} =s_{ab}-{\bar s}_{ab}$ and $p^{ab}=\pi ^{ab}- {\bar \pi}^{ab}$ be
   the differences between the  metric and  momenta and their asymptotic
   AdS values. Let $\xi ^a$ be a Killing field of AdS, and hence an asymptotic Killing vector of $g_{ab}$, which can be decomposed  as
\begin{equation}\label{kvcomp}
\xi^a = Fn^a +\beta ^a \  , \quad n_c\beta^c =0.
\end{equation}
The ADM gravitational charge associated with $\xi ^a$ is then defined by an integral over the
 $(D-2)$-dimensional boundary of $\cal{S}$ in the asymptotic region,
\begin{equation}\label{charge}
Q(\xi )= -{1\over 16 \pi }\int_{\partial\cal{S}_{\infty}} da_c B^c 
\end{equation}
where the boundary vector $B^a$ is given by
\begin{align}\label{bvecdef}
B^a=F({D}^ah-{D}_bh^{ab})-h{D}^aF+h^{ab}{D}_bF+
{1 \over \sqrt{{s}}}\beta^b({\bar \pi}^{cd}h_{cd}{\bar s}^a{}_b-2{\bar \pi}^{ac}h_{bc}-2p^a{}_b)
\end{align}
and $D_a$ is the derivative operator compatible with the background AdS metric $\bar{s}_{ab}$.

The ADM mass $\cal{M}$ is the conserved charge associated with the time-translation symmetry of the asymptotic AdS spacetime. To obtain the explicit expression 
for the mass, one takes $\cal{S}$ to be a constant time slice and  the Killing field in (\ref{charge}) to be 
 $\xi^a = \left(\partial /\partial t\right)^a $.  In terms of the coordinates in (\ref{adsfalloff}), the asymptotic behavior of the normal form is given by  $n_a=-(r/l) \nabla_a t$, while the area element is
$da_c= d^{D-2}x (r/l)^{D-3} \nabla_c r $.
From the definition (\ref{kvcomp}) we have that $F=  r/l$.  Consequently, 
unlike in asymptotically flat spacetimes where
 $F$ is a constant, the terms in (\ref{bvecdef}) with derivatives acting on $F$  contribute to the mass in AdS.  The components of $h_{ab}$ can be read off from equation (\ref{adsfalloff}), and 
 recalling that with this choice of foliation $h_{tt}\equiv 0$, one finds that the ADM mass is given by
\begin{equation}\label{massdef}
 \mathcal{M} ={\vbar \over 16\pi\, l^D }
\left( (D-2)c_r+(D-1)\sum_{i=1}^{D-2}{c_i}\right) ,
\end{equation}
where  $\vbar =\prod _k L_k $ is the volume of the planar box  at infinity.


The tension $\mathcal{T}_i$ is the ADM charge associated with the spatial translation symmetry of the asymptotic AdS metric in the 
$x^i$-direction \cite{Traschen:2001pb,Townsend:2001rg}, and hence the 
Killing vector in (\ref{charge})  is taken to be $\xi^a =\left(\partial / \partial x^i\right)^a $. 
The computation of $\mathcal{T}_i$ is similar to that above for the mass, except that in this case the foliation is taken to be a family of constant spatial coordinate $x^i$.
The boundary integral  (\ref{charge}) then includes a trivial integration over time,
which we divide out, so that $\mathcal{T}_i$ is properly a tension per unit time.  
The boundary integral (\ref{bvecdef}) then gives
\begin{equation}\label{tendef}
L_j \mathcal{T}_j ={\vbar \over 16\pi\, l^D}  \left((D-2)c_r +(D-1)[\sum_{i\neq j}^{D-2}c_i- c_t ]
\right)
\end{equation}
with no sum on $j$ implied on the left hand side.


\subsection{Planar black hole and AdS soliton}\label{examples}

In this section, we use the expressions obtained above to compute the ADM mass and tensions for the planar AdS black hole and AdS soliton \cite{Horowitz:1998ha}.
The planar black hole metric is given by
\begin{align}\label{bhmetric}
ds^2=-\frac{r^2}{l^2}\left(1-\frac{r^{D-1}_0}{r^{D-1}}\right)dt^2+\frac{r^2}{l^2}
\sum_{i=1}^{D-2} (dx^i)^2+\frac{l^2}{r^2}\left(1-\frac{r^{D-1}_0}{r^{D-1}}\right)^{-1}dr^2
\end{align}
with the horizon being  located at $r =r_0$. Substitution into equations 
(\ref{massdef}) and (\ref{tendef}) yields for the planar black hole
\begin{align}\label{bhcharges}
\mathcal{M}=   {(D-2)\vbar\, r^{D-1}_0  \over 16\pi  l^{D} } , \qquad  L_i \mathcal{T}_i=
- {\vbar \, r^{D-1}_0 \over  16\pi l^{D} } 
\end{align} 
where $ i=1,\dots ,D-2$.
The AdS soliton \cite{Horowitz:1998ha}  is obtained from the black hole metric above via a double analytic continuation in time and one of the planar coordinates, yielding the metric   
\begin{align}\label{solitonmetric}
ds^2=
\frac{r^2}{l^2}\left( -dt^2  +  \sum_{i=1}^{D-3} (dx^i)^2   +
 \left(1-\frac{r^{D-1}_0}{r^{D-1}}\right) dz^2 \right)+
\frac{l^2}{r^2}\left(1-\frac{r^{D-1}_0}{r^{D-1}}\right)^{-1}dr^2
\end{align}
Instead of a horizon, the spacelike Killing field $(\partial/  \partial z)$ has a fixed point
at $r=r_0$. To avoid a conical singularity at the fixed point, the coordinate $z$ must be identified with period
 $L_z=4\pi l^2/(D-1)r_0$ \cite{Horowitz:1998ha}. The structure and thermodynamics
 of such bubbles, as well as spacetimes containing combinations of bubbles and black holes,
 have been studied extensively in the $\Lambda =0$ case 
   \cite{Elvang:2004iz,Elvang:2002br,Kastor:2008wd}. Using
 (\ref{massdef}) and (\ref{tendef}) one finds that the gravitational charges for the AdS soliton are given by
\begin{equation}\label{solitoncharges}
L_z \mathcal{T}_z= {(D-2)\vbar \, r^{D-1}_0\over 16\pi \, l^D}, \qquad
\mathcal{M}= L_i \mathcal{T}_i=  -\frac{\vbar\, r^{D-1}_0 }{16\pi \, l^{D}}
 \end{equation}
 where now $i= 1,\dots , D-3$.
One sees that the set of charges for the black hole and for the soliton are related
via  interchanging  $\mass $ and $L_z \ten _z$. In particular, the soliton has
 a negative mass but positive  tension in the $z$-direction, while the black hole has positive
 mass and all tensions negative. In the next section we will show that
 the sum of the mass and the tensions is always zero for solutions, so some of these
 quantities, in fact, needed to have opposite signs. 


\section{Technical matters}\label{trace}
\subsection{Trace condition and Komar expressions}\label{technical}

We now have the ADM expressions for the mass and tensions.  However, the Smarr formulas result from applications of Komar integrals that naturally yield the equivalent Komar expressions for these quantities,  the two sets of expressions being related via the field equations.  This is familiar from the asymptotically flat case, where the ADM expression for the mass involves the fall-off coefficient $c_r$, while the Komar expression depends instead on $c_t$.  The field equations in the asymptotic regime relate these two coefficients and yield the equivalence between the ADM and Komar expressions for the mass.  We need to find the analogous relations, involving the fall-off coefficients $c_i$ as well, with our planar AdS boundary conditions (\ref{adsfalloff}).

In order to do this, it is sufficient to expand the trace of the field equations to first order about the AdS metric in the asymptotic region.  Because the cosmological term in (\ref{einstein}) does not contribute to the trace, we are looking simply at the linearization of $R=0$ around AdS.  Writing $g_{ab}=\bar g_{ab}+\gamma_{ab}$, with $\bar g_{ab}$ the AdS metric, we then have
\begin{align}\label{deltar}
 - \bnabla_a \bnabla^a \gamma + \bnabla_a \bnabla_b \gamma^{ab} - \bar{R}_{ab}\gamma^{ab}=0
\end{align} 
where $\bnabla_a$ is the AdS covariant derivative operator, indices are raised and lowered with the AdS metric and  $\bar R_{ab}=2\Lambda\bar g_{ab}/(D-2)$.
Upon substitution of the asymptotic form of the metric (\ref{adsfalloff}) into this equation, 
each of the first two terms on the left hand side vanishes identically, reducing the equation to $\Lambda \bar{g}^{ab}\gamma_{ab}=0$.
In terms of the fall-off coefficients this implies that
\begin{equation}\label{cocon}
-c_t  +  c_r + \sum_{i=1}^{D-2} c_i  =0 .
\end{equation}
 One readily checks that the planar black hole and AdS soliton solutions in Section (\ref{examples}) satisfy this condition\footnote{A second, instructive way to derive equation (\ref{cocon}) is in terms of the far field limit of the Killing potentials introduced in Section (\ref{smarrsection}) below.  This alternative demonstration of (\ref{cocon}) is presented in Appendix A.}.

With the relation (\ref{cocon}) in hand, we can eliminate the fall-off coefficient $c_r$ from the expressions for the
 $(D-1)$ charges in  (\ref{massdef}) and (\ref{tendef}) yielding the Komar-type expressions
\begin{eqnarray}\label{massandten}
M &=& {\vbar \over 16\pi\,  l^D} \left( (D-2) c_t + \sum_ic_i  \right)  \\ \label{massandten2}
L_j \ten_j  &=& {\vbar \over 16\pi\,  l^D}  \left(- (D-2) c_j  + \sum_{i\neq j } c_i -c_t  \right) 
 \end{eqnarray}
with $i, j=1,\dots , D-2$.
From these expressions one can read off an important property of asymptotically planar AdS spacetimes, namely that 
the sum of the mass plus tensions of solutions necessarily  vanishes for such spacetimes, 
\begin{equation}\label{zerotrace}
\mathcal{M}+  \sum_i L_i \mathcal{T}_i =0.
\end{equation}
Equation (\ref{zerotrace}), which is not readily apparent from formulas (\ref{massdef}) and  (\ref{tendef}), 
can be viewed as a kind of  Smarr relation that holds on solutions with planar AdS asymptotics, although non-standard in the sense that it relates only the  ADM
charges, without involving the interior geometry of the spacetime.  Note that if we regard the mass and tensions as the diagonal elements of an ADM stress tensor in the planar directions, then (\ref{zerotrace}) has the form of a zero trace condition.

This relation is imposed by the AdS asymptotics.
It has  no analogue, for example, in Kaluza-Klein theory with $\Lambda=0$, where the mass and tensions are independent quantities \cite{Elvang:2004iz,Kastor:2006ti}.  
In contrast, we see that in AdS the mass and tensions are not independent.
We expect that the zero trace condition (\ref{zerotrace}) should in some way be a consequence of  the  conformal isometry of AdS 
generated by the Killing field $V^a = \sum_\alpha x^\alpha (\partial / \partial x^\alpha ) -r (\partial / \partial r )$.
However this
is not apparent from the calculation given here and will be pursued in future work.
    
\subsection{Gauge fixing}\label{gauge}    
The task at hand is now to invert the formulas in (\ref{massandten}) and (\ref{massandten2}) to
 give expressions for the fall-off coefficients $c_t$ and $c_i$ in terms of the ADM mass and tensions.
If one regards these formulas as a single vector equation relating the vector of charges $(M,\ten_i)$ in terms of the vector of fall-off coefficients $(c_t,c_i)$, then one finds that the matrix has vanishing determinant, reflecting the fact that the charges must end up satisfying the constraint (\ref{zerotrace}). 
In order to proceed, one charge must be removed from the vector and considered to be determined in terms of the others through
equation (\ref{zerotrace}).   We will take $\ten _1 $
to be this auxillary tension, and proceed with the reduced vector of $D-2$ charges.  
  
  However, this leaves the counting off.   There are now only $(D-2)$ charges, but there are still
$(D-1)$ coefficients, and therefore the matrix is no longer square. The underlying problem
is that the coordinates have not been completely fixed by the asymptotic form of the metric in equation  (\ref{adsfalloff}). 
One can still perform a  coordinate transformation that maintains the form of (\ref{adsfalloff}). This residual gauge freedom can be fixed  {\it e.g.} by taking one of the 
fall-off coefficients to vanish, an approach that is  similar to employing
synchronous gauge in cosmology. We will take $c_{D-2} =0$,
which is convenient since the metrics given above for the planar black hole (\ref{bhmetric}) and AdS soliton (\ref{solitonmetric}) already have this property. 
This gauge choice may be implemented via the coordinate transformation $x^{\prime a} =x^a +\chi ^a$
where $\chi_r =kr^{-D}$ with $k$ constant, and the remaining components $\chi_\alpha=0$. It follows that $\nabla _r \chi _r =
-(D-1)kr^{-(D+1)}$ and that $\nabla_\alpha \chi_\beta + \nabla_\beta \chi_\alpha =
(2k/ l^4 ) \eta_{\alpha\beta}r^{-D+3}$. Hence, the constant $k$ may be chosen such that the coefficient $c_{D-2}=0$
in the new coordinates, while the other coefficients 
are merely relabeled. As a check, direct substitution shows that the expressions for the ADM charges are invariant under the coordinate transformation. 

The reduced set of equations obtained by excluding the relation
for $L_1 \ten_1$ and setting $c_{D-2}=0$ in (\ref{massandten}) and (\ref{massandten2}) can now be inverted to give
\begin{eqnarray}\label{inversion2}
 c_t  &=& \frac{16 \pi l^D }{(D-1) \vbar }\left( \mathcal{M} -L_{D-2}\ten_{D-2} \right)
 \\\nonumber
 c_i &=&   \frac{16 \pi l^D }{(D-1)\vbar }\left( L_{D-2}\ten_{D-2}
-  L_i \mathcal{T}_i  \right), 
\end{eqnarray}
where $i=1, \dots ,(D-2)$.

\subsection{Comparison with AdS/CFT boundary stress tensor}

In this section we compare the expressions we have found for the ADM gravitational charges of asymptotically planar AdS spacetimes to those computed using the  quasilocal boundary stress tensors method \cite{Brown:1986nw,Balasubramanian:1999re,Myers:1999psa}.
 From the point of view of AdS/CFT one expects that, in the absence of  a conformal anomaly, the boundary stress-energy tensor
 should be traceless.
 For definiteness, let us consider the five-dimensional case
where the boundary stress tensor $\tau_{\alpha\beta}$ is given by  \cite{Balasubramanian:1999re}
\begin{equation}\label{stress}
8 \pi  \tau_{\alpha\beta}=\Theta_{\alpha\beta}-\left(\Theta+\frac{3}{l}+\frac{l}{4}R\right)h_{\alpha\beta}-\frac{l}{2}R_{\alpha\beta},
\end{equation}
where $h_{\alpha\beta}$ is the induced metric on a  hypersurface of constant $r$
in the asymptotic region, $\Theta_{\alpha\beta}$ is the extrinsic curvature of
this slice, and $R_{\alpha\beta}$ is the Ricci tensor for the metric $h_{\alpha\beta}$.  
 The leading contributions to
 $R_{\alpha\beta}$ arise, in principle, from second order terms in an expansion around infinity.
However, these contributions vanish
for the planar geometry \cite{Balasubramanian:1999re}.
Using the asymptotic form of the  metric in (\ref{adsfalloff}) 
the extrinsic curvature components are found to be
\begin{align}
\Theta_{tt}=\frac{r^2}{l^3}+\frac{1}{l^3r^2}\left( c_t -\frac{c_r}{2} \right),
 \qquad \Theta_{ij}= \left( -\frac{r^2}{l^3}+\frac{1}{l^3r^2}\left( 1+\frac{c_r}{2} \right) \right) \delta_{ij}
\end{align}
These terms include divergent contributions from the AdS background
that are cancelled by the  $3h_{\alpha\beta}/l$ term in (\ref{stress}).
The components of the boundary stress tensor are then given by
\begin{align}
\tau_{tt}=\frac{1}{16\pi\, l^3r^2}\left( 3c_r   +4 \sum_i c_i \right), 
\qquad   \tau_{i j}=\frac{1}{16\pi\, l^3 r^2}\left(  4c_t-4 \sum_{k\neq i} c_k  -
3c_r  \right) \delta_{ij}
\end{align}
These match respectively with the integrands that gave rise to 
the expressions for the ADM mass and tensions given in (\ref{massdef}) and
(\ref{tendef}) for $D=5$.
The boundary stress tensor charges are  defined for Killing fields $k^a$ of the 
boundary metric. Decomposing the $D-1$ dimensional metric according to
\begin{align}
h_{ab}=(n \cdot n) n_{a}n_{b}+\sigma_{ab} \quad n \cdot n = \pm 1 \quad n^{a}\sigma_{ab}=0
\end{align}
with $\sigma_{ab}$ the metric on a $(D-2)$ dimensional surface contained in the boundary.
The charge
 is then defined as an integral over a $(D-2)$ dimensional surface
\begin{equation}\label{stresscharge}
Q(k )_{bdry} =\int_{\partial\Sigma} d^{D-2}x \sqrt{\sigma}\left(n^{a}\tau_{ab}k^{b}\right)
\end{equation}
Since  time and space translation Killing vectors of AdS are also Killing vectors of
the boundary, this shows that  the charges in the boundary 
formalism match the ADM quantities. 

\section{ Deriving the Smarr formulas}\label{smarrsection}

In this section, we derive  Smarr formulas for  static asymptotically planar AdS spacetimes  
having $0\leq p\leq (D-2)$ spacelike Killing vectors that act as spatial translations at infinity, and which may contain black holes and/or bubbles.  The first step is to write down a Komar integral identity associated with each Killing field,  using the generalization of this procedure to $\Lambda\neq 0$ given in \cite{Kastor:2008xb}.   
These Komar
   identities will relate the asymptotic behavior of the spacetime (\ref{adsfalloff}),
  to geometric properties of the horizon or bubble. The second step is to use  
  equation (\ref{inversion2}) 
 to convert the Komar identities into statements in terms of $\mass$ and $\ten_i$.
  A related construction has appeared in reference
 \cite{Papadimitriou:2005ii}, but focused only on the static Killing field.

Let $g_{ab}$ be a metric satisfying the Einstein equation with cosmological constant
(\ref{einstein})
and let $\xi ^a$ be a Killing field of this metric.   The generalization of the Komar integral identities to $\Lambda\neq 0$ in \cite{Kastor:2008xb} involves
an antisymmetric Killing potentials, satisfying
\begin{equation}\label{kp}
\nabla_a \omega^{ab}=\xi^b ,
\end{equation}
which necessarily exist because $\nabla_a\xi^a=0$.
Killing potentials are not unique.  A solution to the homogeneous equation can always be added.  However, this non-uniqueness does not affect the result, which comes about as follows.  One starts from
the differential identity  $\nabla^a \nabla_a \xi^b=-R^b{}_a \xi^a$,
which  on solutions to the field equations can be rewritten as  $\nabla^a \nabla_a \xi^b=-2\Lambda \nabla_c \omega ^{cb}$.
The  Killing potential allows  the right hand side to be cast as a total divergence, so that
Stokes theorem can be used.
Let  the spacetime be foliated  by surfaces ${\cal S}$ with unit normal $n_b$
 as in (\ref{foliation}), and let $\Sigma$
 to be a volume contained in ${\cal S}$ with boundaries $\partial \Sigma_I$. 
 Integrating the differential relation over $\Sigma$ then gives the Komar integral identity 
 \cite{Kastor:2008xb}
\begin{equation}\label{komar}
\sum _I \int_{\partial \Sigma_I} ds_{ab}
\left(\nabla^a \xi^b +\frac{2\Lambda}{D-2}\omega^{ab} \right)=0 
\end{equation}
where $ds_{ab} = \rho_{[a} n_{b]}\, da$ with $\rho _a$ being the outward pointing unit normal
to  $\partial \Sigma_I$ within ${\cal S}$.
There can be boundary contributions at infinity, black hole horizons, and bubbles, so that the Komar integral naturally splits up as
 \begin{equation}\label{komartwo}
  I_\infty + I_H +I_{bub} =0
 \end{equation}
 where each of the terms corresponds to the integrals in (\ref{komar}) over a given type of boundary component.
 We proceed with this construction first for the static Killing vector and then for the 
 spatial translation Killing fields. The derivation assumes that the black hole has
 a bifurcate horizon and that the spacelike Killing fields are tangent to the horizon.


\subsection{Komar identity for the static Killing field}

Let the Killing field $\xi^a$ in the Komar integral identity (\ref{komar})  be the static
Killing field, $\mathcal{S}$ be a constant time slice which intersects the horizon
at the bifurcation surface, and first consider the boundary integral at infinity. One finds that some individual terms in this are 
divergent, but that these divergences cancel when the full integrand in (\ref{komar}) is assembled\footnote{Note that this has to be the case, if the integrals
on the horizon and bubble are finite, since the Komar integral identity requires that these terms sum to zero.}.   This calculation proceeds as follows.
Near infinity, one has 
$\rho_a n_b =- \nabla _a r \nabla _b t $ and
the area element in (\ref{komar}) is therefore given by
 $da=(r/l )^{D-2}\, d^{D-2}x $. Finite contributions to the integrals will then come from terms that 
fall off as $r^{-(D-2)}$.  One finds that
 \begin{equation}\label{onet}
 \rho_a n_b \,  \nabla^a \xi^b =
 - {r\over l^2 } - {1\over 2l^2 r^{D-2}}  \left((D-2)c_t -2 c_r \right)
\end{equation} 
The first term on the right hand side will give a divergent contribution to the integral.
This is cancelled \cite{Kastor:2009wy} by a contribution from the 
Killing potential $\omega^{ab}_{(t)}$, where the label $(t)$  indicates the corresponding Killing vector, \cite{Kastor:2009wy}. The 
Killing potential may be written in the form $\omega^{ab}_{(t)}  =  \omega^{ab}_{(t)AdS}+  \Delta \omega^{ab}_{(t)}$
where $\omega^{ab}_{(t)AdS}$ is the Killing potential in the background AdS spacetime, satisfying 
\begin{align}
{\bnabla}_a \omega^{ab}_{(t)AdS}={\xi}^b 
\end{align}
This can be taken to be  $\omega^{rt}_{(t)AdS}=r/(D-1)$
and further recalling that $2\Lambda /(D-1)(D-2) =-1/l^2$ one has for the Killing potential term in (\ref{komar})
\begin{align}\label{twot}
\frac{2\Lambda}{D-2}{\rho}_a {n}_b \omega^{ab}_{(t)} =
\frac{r}{l^2} + \frac{1}{2l^2 r^{D-2}}\left( -c_t  + c_r\right) + \frac{2\Lambda}{D-2} {\rho}_a {n}_b\Delta \omega^{ab}_{(t)}
\end{align}
We see that the first term on the right hand side will cancel with the corresponding term in (\ref{onet}) giving an overall finite result for the boundary integral at infinity
\begin{align}
 I_{\infty}= \frac{\vbar }{2 l^D}\left(2c_r-(D-1)c_t\right)
+{2\Lambda\over D-2} \int_{\partial\Sigma_{\infty}} ds_{ab} \Delta \omega^{ab}_{(t)}
\end{align}
where again the constant $\vbar = \prod_k L_k$.

We next turn to the horizon boundary term.  Evaluation of the $\nabla_a\xi_b$ term at the horizon proceeds as in the $\Lambda =0$ case \cite{Bardeen:1973gs}, 
and is proportional to the surface gravity and combining this with the Killing potential term gives
\begin{align}
I_H=  \kappa_H A_H - \frac{2\Lambda}{D-2}\int_{\Sigma_H}ds_{ab}\omega^{ab}_{(t)} 
\end{align}
The net contribution to (\ref{komar}) from the Killing potential will be the difference between its integrals 
at infinity and at the horizon, and
it becomes useful to define quantities  $\Theta_{(J)}$ for each Killing vector $\xi^a _{(J)}$,  where $J=t,1,\dots,p$, in terms of these differences by
\begin{align}\label{thetadef}
\Theta_{(J)}=  \sum_{in} \int_{\Sigma_{in}}ds_{ab} \omega_{(J)}^{ab} -\int_{\Sigma_{\infty}}ds_{ab} \Delta \omega_{(J)}^{ab}
\end{align}
where the sum is over interior boundaries, including both black hole horizons and bubbles.
For the static Killing field
there is no boundary contribution at the bubble, so long as it is smooth, and the Komar 
identity is then given by $I_\infty + I_H =0$.  Assembling the various ingredients then yields
\begin{eqnarray}\label{smarrtime}
{ \kappa_H A_H \over 8\pi} -  \frac{\Theta_{(t)}   \Lambda }{ 4\pi (D-2) }   & = &
  \frac{\vbar }{16\pi  l^D}\left((D-1)c_t-2c_r\right) \\  \nonumber 
&=& -  \sum _{i=1} ^{D-3} L_i \ten _i 
\end{eqnarray}
where the second equality follows from using equations (\ref{cocon}) and (\ref{inversion2}) to replace the fall-off coefficients with the gravitational charges. The fact that the sum over tensions
only goes to $i=D-3$ rather than $i=D-2$ is due to the choice of gauge $c_{D-2} =0$;
examining equations (\ref{inversion2}) shows that $\ten_{D-2}$ drops out of the
linear combination of the fall-off coefficients that appear in (\ref{smarrtime}).
 The final Smarr
formulas will not have this apparent gauge dependence, with one proviso to be noted below. When the spacetime
is static but has no spatial Killing fields there is one puzzle, which we discuss
at the end of Section (\ref{results}).

\subsection{Komar identities for spatial translations without fixed points}

Now we turn to the Komar integral identities associated with the $p$ spacelike Killing vectors, which we divide into two cases depending on
whether the Killing field has a fixed point, or not. 
First consider the case when the Killing field
$X^a =(\partial / \partial x )^a $ does not have a fixed point, and let
the spacetime be foliated by slices $\mathcal{S}_x$ of constant coordinate $x$, which are
Lorentzian. 
Given the assumed absence of fixed points, there are again boundary terms only at infinity and at the black hole horizon.
Each of these terms contains an integral over the time direction, which because the spacetime is assumed to be static, contributing only an overall factor of a time interval 
$\Delta t$. 
The evaluation of the boundary term at infinity is very similar to the calculation for the 
static Killing field.
Near infinity ${\rho}_a {x}_b = \nabla _a r\nabla_b x $ and the area element is $ da=(r/l )^{D-2}\, dt dx^{D-3}$ and as before the divergent term that arises from the Killing vector piece is canceled by the asymptotically AdS behavior of the Killing potential 
$\omega^{ab}_{(x)} $. Dividing through by the common factor of $\Delta t$ and 
 multiplying  through by the length $L_x$, we find that the boundary term at infinity is given by
\begin{align}\label{spinf}
 \left({L_x \over \Delta t} \right) I_{\infty} =-\frac{\vbar }{2 l^D}((D-1)c_x+2c_r)+
\frac{2\Lambda}{D-2} \left({L_x \over \Delta t} \right)
\int_{\partial \Sigma_{\infty}}ds_{ab} \Delta \omega^{ab}_{(x)}
\end{align}

In order to evaluate the boundary term at the horizon, we choose the slice  such that  the unit normal ${x}_a$ 
is proportional to the Killing field, so that $X^a =F  {x}^a$.
The boundary contribution  from the derivative of the spatial
Killing field in (\ref{komar})  can then be seen to vanish on the horizon, as follows.  
The generator  $\xi^a$ is normal to the horizon, so that the Killing vector term in the integrand of (\ref{komar}) is proportional to
\begin{align}
{x}^b \xi^a \nabla_a X_{b}= - {x}^b \nabla_b (\xi ^a X_a)+{x}^b X^a \nabla_b \xi_a . 
\end{align} 
The first term on the right hand side of this equation vanishes since $X_a \xi ^a =0$ on the horizon, and the derivative is tangent
to  the horizon. The second term vanishes because it is the contraction of an antisymmetric
tensor and a symmetric tensor. 
For these vector fields there is no contribution from the bubble and therefore the Komar identity becomes
\begin{eqnarray}\label{smarrspatial}
- \frac{  \Lambda  \Theta_{(x)} }{4\pi (D-2) }  \left({L_x \over \Delta t } \right)   
 & =& \frac{\vbar }{16 \pi  l^D} \left(  (D-1)c_x +2c_r \right)  \\  \nonumber 
&=&  \mathcal{M} +  \sum_{\substack {i=1 \\ i\neq x}}^{D-3}L_i \mathcal{T}_i 
\end{eqnarray}
where the quantity $\Theta_{(x)}  $ is defined in equation (\ref{thetadef}). 
The Smarr formula (\ref{smarrspatial}) holds for  any spacelike Killing field $(\partial/\partial x)^a$
 with no fixed point.


\subsection{Adding fixed points}
 
 The remaining case is when a spacelike Killing field generates a bubble, that is,
 $Z^a =(\partial / \partial z )^a$  has
 a fixed point on a $(D-2)$-dimensional Lorentzian submanifold $\cal{B}$.
 A bolt in the classification of Gibbons and Hawking \cite{Gibbons:1979},
 we refer to these  as  bubbles,  following the work of \cite{Elvang:2004iz,Elvang:2002br,Kastor:2008wd}. The Killing field
$Z^a$ acts like a spatial translation around a compact dimension at infinity, but
like a rotation in the vicinity of the bubble.
The AdS soliton metric (\ref{solitonmetric}) has a bubble at $r=r_0$, which has the topology of a  $(D-3)$-dimensional spatial torus
cross time. When we refer to the area of a bubble, we will mean the area of the spatial torus at fixed time. The derivative of the Killing field $Z_a$
at the bubble is normal to the bubble, so that
\begin{equation}\label{bubderiv}
\nabla_{[a} Z_{b] }=\kappa_B ( e_a^{(1)} e_b^{(2)}  -e_a^{(2)} e_b^{(1)} )
\end{equation}
where $e_a^{(1)},e_a^{(2)}$ are a mutually orthonormal set of  normal forms to $\cal{B}$
 \cite{Gibbons:1979} and $\kappa _B$ is constant on the bubble \cite{Kastor:2008wd}.
 It follows from (\ref{bubderiv}) that
 $\kappa _B ^2 = (1/2)( \nabla_a Z_b) \nabla ^a Z^b $, which except for the sign is the identical to
 the formula for the surface gravity of a black hole horizon. The AdS
 soliton (\ref{solitonmetric}) has $\kappa _B =(D-1)r_0 /2l^2 $.
 The geometry at the bubble is the same as that of a Euclidean black hole with the constant $\kappa _B$
 playing the role of surface gravity. Smoothness of the Euclidean 
 black hole determines the periodicity in Euclidean time and hence its temperature.
 Smoothness of the bubble geometry requires that the periodicity of the $z$-coordinate be such that $\kappa _B  L_z =2\pi  $. 
 
 We are now prepared to evaluate the Komar integral expression (\ref{komar}) for spacelike Killing vectors having fixed points. The
 boundary terms at the horizon and at infinity are unchanged from the case with no fixed point, with the only new contribution coming from $I_B$.
 Equation (\ref{bubderiv}) implies that $e_a^{(1)} e_b^{(2)} \nabla ^a Z^b =\kappa _B$,
and we arrive at the expression
\begin{equation}
I_B = -A_B \kappa_B\, \Delta t +
{2\Lambda\over D-2} \int_{\partial\Sigma_{\infty}} ds_{ab} \Delta \omega^{ab}_{(z)}.
\end{equation}
where $A_B$ denotes the area of a $(D-3)$-dimensional spatial torus. 
Combining the three boundary terms, multiplying by $L_z$ and dividing out a common
factor of $\Delta t$ gives the Smarr relation for a spacelike Killing field with fixed point
\begin{eqnarray}\label{smarrbubble}
  -\frac{\kappa_B L_z A_B}{8\pi } 
   -  \frac{ \Lambda  \Theta_{(z)} }{4 \pi (D-2) }  \left({L_z \over \Delta t } \right)  &=&
  \frac{\vbar }{16 \pi  l^D}((D-1)c_z +2c_r) \\  \nonumber \cr
  &=& \mathcal{M} +  \sum_{\substack {i=1 \\ i\neq z}}^{D-3}L_i \mathcal{T}_i  
\end{eqnarray}
where $\Theta_{(z)}  $ is defined in (\ref{thetadef}). 
Although we are primarily interested in smooth bubbles, the derivation allows for a possible angular deficit  $\psi= 2\pi -\kappa_B L_z$ at the bubble \cite{Kastor:2008wd}.
As long as $0< \psi \leq 1$ this is a conical singularity with positive integrated
curvature and corresponds to a $(D-3)$-brane wrapping the bubble.

\section{Organizing the results}\label{results}

The purpose of this section is to rewrite the set of Smarr relations (\ref{smarrtime}),
 (\ref{smarrspatial}) and
(\ref{smarrbubble}) in a more transparent form. We start by finding conditions such that
the different thermodynamic potentials $\Theta _{(J)}$ are all proportional to a common thermodynamic volume $V$.

\subsection{Thermodynamic potentials and volumes}
It was shown in \cite{Kastor:2009wy} that the quantity $\Theta _{(t)}$ appearing in the Smarr relation (\ref{smarrtime})  has a simple interpretation as minus the volume occupied by the black hole.  A similar interpretation may be applied to all the potentials $\Theta _{(J)}$ appearing in (\ref{smarrspatial}) and
(\ref{smarrbubble}).  The boundary integral expression for 
 $\Theta _{(J)}$ in (\ref{thetadef})
can be converted to a volume integral by means of Gauss's law. Choose the 
slice connecting the relevant boundaries such that its normal is in the direction of the Killing field,
{\it i.e.} so that $\xi^a _{(J)} = F_{(J)} n^a _{(J)}$.  Making use of the decompositions
(\ref{foliation}) and (\ref{kvcomp}), it then follows that for each $J$ one has
 $\int_{\partial\Sigma}\rho_a n_b\, \omega ^{ab} =
n\cdot n \int _\Sigma F \sqrt{|s|}$.
Since the integrand at infinity in (\ref{thetadef}) depends on
the difference between the Killing potentials for the metric $g_{ab}$ and for the AdS metric $\bar g_{ab}$, the volume expression for 
$\Theta_{(J)}$ is likewise the difference between two integrals
 \begin{equation}\label{thetavol}
 \Theta _{(J)} = n_{(J)}\cdot n_{(J)} \left[ \int_{\Sigma _{J,AdS}} F_{(J,AdS)} \sqrt{ | s _{J,AdS} |} 
 - \int_{\Sigma_{(J)} } F_{(J)}\sqrt{| \ s_{(J)} |} \right] 
\end{equation}
Here, the pure AdS volume integral extends inward all the way to $r=0$, while the other volume integral extends only down to a black hole horizon or soliton bubble radius. 
For this reason, it was argued in \cite{Kastor:2009wy,Dolan:2010zz} that the quantity $-\Theta_{(t)}$ may be thought of as  the volume occupied by the black hole. 
The properties of this thermodynamic volume were further explored in
 \cite{Cvetic:2010jb} where it was 
found that for rotating black holes the thermodynamic volume differs from the
geometric volume in such a way that suggests a `reverse isoperimetric inequality' for AdS black holes  \cite{Cvetic:2010jb}.

When the index $J$ in (\ref{thetavol}) refers to one of the spatial Killing vectors, one has $n\cdot n = +1$ and the potential
$ \Theta _{(i)} $ is positive, with the quantity $L_i  \Theta _{(i)} /\Delta t$ corresponding to a $(D-1)$-dimensional
spatial volume. It is useful to define in each of these cases the positive volumes $V_{(J)}$ according to
\begin{equation}\label{voldefs}
V_{(t)}   = - \Theta _{(t)},  \qquad   V_{(i)} =  \left({L_i \over \Delta t} \right)  \Theta _{(i)}
 \end{equation} 
 with $ i=1,\dots, p$.
 For the planar black hole and the AdS soliton spacetimes these volumes are redundant
 quantities. Due to symmetries and the diagonal form  of the metric
 all the $V_{(J)}$ are equal to a common volume $V$, 
 \begin{equation}\label{equalvol}
   V_{(J)} =V
 \end{equation}
 for $J=t, 1,\dots,p$, where in these cases we have $p=(D-2)$.
This raises the question as to how many independent volumes there are in general? 
We will show next that if two of the Killing vectors commute, then their associated volumes are
necessarily equal. Therefore, a sufficient condition that all the thermodynamic volumes $V_{(J)}$ in the
Smarr relations are equal to a common volume is that the set of Killing vectors be mutually 
commuting.

To demonstrate this, we need to show that if two of the Killing vectors commute, that the quantity $ F_{(J)}\sqrt{| \ s_{(J)} |}$ will be the same for both vectors.
So, assume that $\xi^a $ and $\eta^a $ are two commuting Killing vectors of an asymptotically planar AdS spacetime, with
$\xi\simeq  \partial/ \partial t$ and  $\eta \simeq \partial / \partial y$
in the asymptotic region. The area elements spanned by these vectors are then integrable and
the metric can be put in block diagonal form (see {\it e.g.} the discussion in
\cite{Carter:1987})
\begin{equation}\label{blockmetric}
ds^2=g_{tt} dt^2 + 2g_{ty} dt dy   + g_{yy} dy^2
+ \sum_{i,j =2}^{D-1}q_{ij} dx^i dx^j
 \end{equation}
 where $q_{ij} = g_{ij} $.
  Now consider  a constant time slice with normal $n_a = -F_{(t)} \nabla _a t$. 
If we decompose the Killing vector $\xi^a$ according to (\ref{kvcomp}), one has $\xi ^a = F_{(t)} n^a + \beta ^a$ where  $g_{ty}= \beta _y$ and
$g_{tt} = -(F_{(t)} )^2 + \beta ^y \beta_y$. This allows us to write  the volume element for (\ref{blockmetric}) as
$\sqrt{-g} =F_{(t)} \sqrt{g_{yy} } \sqrt{q}$.  One also has $\sqrt{s_{(t)}} =\sqrt{g_{yy} } \sqrt{q}$, which results in the statement $ \sqrt{-g} =F_{(t)}  \sqrt{s_{(t)}}$.
These steps may be repeated for a slice of constant $y$, with the second Killing vector $\eta ^a$ decomposed to find that 
  $\sqrt{-g} =F_{(y)} \sqrt{- g_{tt} } \sqrt{q} =
 F_{(y)}  \sqrt{-s_{(y)} }$.
Comparing the expressions for the volume elements from these two slicings establishes that  $ F_{(t)}\sqrt{| \ s_{(t)} |} = F_{(y)}\sqrt{| \ s_{(y)} |}$, which implies in turn that
$\Theta_{(t)} =-(L_y /\Delta t) \Theta_{(y)} $ or equivalently  the equality of the two thermodynamic volumes, $V_{(t)} =V_{(y)} $. Extending this argument
to a larger set of commuting Killing vectors implies that the associated thermodynamic volumes
will be equal.

\subsection{Commuting Killing Fields}\label{resultstwo}
Equations (\ref{smarrtime}), (\ref{smarrspatial}) and (\ref{smarrbubble}) give
the Smarr relations that result from the static and spatial translation Killing fields. 
In this section we process these formulas to make their physical implications
more apparent. We will  assume that the spacetime
has at least one freely acting spatial translation Killing field  and that all the Killing fields commute,
so that all the thermodynamic volumes are equal.  At the end of the section we will briefly discuss
the cases in which either the Killing fields
do not commute, or there are no spatial symmetries. 

 For each freely acting spatial translation Killing field, 
 the sum of equations (\ref{smarrtime}) and (\ref{smarrspatial}) gives
\begin{equation}\label{spatialtwo}
M-   L_{F_j} \ten _{F_j} =
{\kappa_H A_H \over 8\pi}  
\end{equation}
 with $ j=1,\dots, n_F$ with $n_F$ being the total number of freely acting spatial Killing vectors.
 Equation (\ref{spatialtwo}) implies
 that  the product $ L_{F_j} \ten _{F_j} $ is the same for all the freely acting  Killing vectors,
 {\it i.e.} that 
\begin{equation}\label{sameten}
 L_{F_j} \ten _{F_j}  = (L\ten)_F 
 \end{equation}
If there are also $n_b $ spatial Killing vectors that having fixed points at a bubble, where necessarilly $n_b +n_F =p $,  then the sum of equations (\ref{smarrtime}) and (\ref{smarrbubble}) gives for each $B_i  $ 
\begin{equation}\label{bubbletwo}
M-  L_{B_i} \ten _{B_i}   =  {\kappa_H A_H \over 8\pi}  
 - { \kappa_{B_i}  L_{B_i} A_{B_i} \over 8\pi} 
 \end{equation}
 where $i= 1,..., n_b$
and  $A_{B_i}$ is the total area of bubbles\footnote{Multi-bubble solutions 
with $\Lambda =0$ are presented in \cite{Elvang:2004iz}, although analogous
solutions with $\Lambda \neq 0$ are not known at present.} corresponding to zeroes of the Killing vector
$\partial/ \partial x^i $. 
Recall that a smooth bubble has $\kappa_{B_i} L_{B_i} = 2\pi$. 

In order to process the Smarr formula (\ref{smarrtime}), we note that the left hand side is equal to
$\mass + L_{D-2}\ten_{D-2}  $. If we now assume that 
 $\partial / \partial x^{D-2}$ is a  freely acting Killing 
field\footnote{It appears to be inconsistent for $\partial / \partial x^{D-2}$ to be non-freely acting.  In Section (\ref{gauge}) we showed that
it is always possible via a coordinate transformation in the asymptotic region to set one of the falloff  coefficients, taken to be $c_{D-2}$, equal to zero. Such a coordinate transformation necessarily changes the components of the metric in the
interior as well. If  $\partial / \partial x^{D-2} $ were not freely acting, 
then both the metric and its normal derivatives would be fixed at the corresponding bubble. Hence trying
to specify the behavior both at the bubble and at infinity will in general not be possible, as
can be checked for the AdS soliton metric (\ref{solitoncharges}).}, recalling that by previous assumption there is at least one such Killing field, then it follows that $L_{D-2}\ten_{D-2} =(L\ten)_F  $
as in  equation  (\ref{spatialtwo}). Substitution for $L_{D-2}\ten_{D-2}$ then yields the Smarr formula
\begin{equation}\label{massresult}
\mass  =  { \kappa_H A_H \over 8\pi} + \frac{   \Lambda  V }{ 8\pi (D-2) } 
\end{equation}
given in the introduction.
These Smarr formulas  give geometrical insight into
the values of the mass and tensions for the planar black hole and AdS soliton solutions given in equations (\ref{bhcharges}) and (\ref{solitoncharges}). 
We see that  $\ten_F$ is always negative (see (\ref{introsmarr})) 
and that $\mass$ is always greater than
or equal to $\ten_F$, because they  differ by the horizon area.
Substituting (\ref{spatialtwo}),  (\ref{bubbletwo}), (\ref{massresult})
into (\ref{smarrtime}) and  using the trace constraint additionally gives a formula for the sum of the $D-2-p$ tensions
not associated with symmetry directions
\begin{equation}\label{nonsymm}
-\sum_{not\, KV} L_i \ten_i   =   \sum_{i=1}^{n_b} { L_{B_i}  \kappa_{B_i} A_{B_i} \over 8\pi}  
 +  { \kappa_H  A_H \over 8\pi} + 
\frac{   (p+1)  \Lambda V}{ 8\pi (D-2) } .
 \end{equation}
To summarize, the relations
  (\ref{spatialtwo}), (\ref{bubbletwo}), (\ref{massresult}), and (\ref{nonsymm})
 hold for spacetimes that are static, with $p$ spatial-translation Killing
fields, at least one of which is freely acting, and assuming that all the Killing fields
are mutually commuting. These Smarr formulas along with the zero trace constraint 
(\ref{zerotrace}) are the main results of this paper.

The  symmetry
between bubbles and black holes noted in the introduction is apparent here.  The equations are invariant
if $\mass$ and $ \kappa _H A_H $ are interchanged with $\ten_{B_i}$ and 
$ \kappa_B  L_{B_i} A_{B_i} $. In the AdS/CFT correspondence,
 the black hole corresponds to a high temperature, deconfined phase of the gauge 
 theory, while the soliton corresponds to the low temperature confining phase
  \cite{Witten:1998zw,Witten:1998qj,Horowitz:1998ha,Surya:2001vj}.   It would be interesting to understand the nature of the symmetry between these phases that is suggested by our results.
  One intriguing possibility allowed by the equations is a critical, or transition, spacetime 
 containing a mix of bubbles and black holes such that  $M=\ten_B =0$. In this case, the trace constraint
 would have to be satisfied by balancing the negative  $\ten_F$ with positive 
 non-symmetry $\ten_i$.

\subsection{More general case}
When the Killing fields do not commute we need to allow for the possibility that
the thermodynamic volumes are not equal. 
Repeating the steps of the previous subsection gives
\begin{equation}\label{spatialnc}
M-   L_{F_j} \ten _{F_j} =
{\kappa_H A_H \over 8\pi}  +  \frac{   \Lambda (V_{(t)} -V_{(F_j)} ) }{ 4\pi (D-2) } , 
\end{equation}
where $ j=1,\dots, n_F$ and additionally with $ i= 1,..., n_b$
\begin{equation}\label{bubblenc}
  L_{B_i} \ten _{B_i}  - L_{F_j} \ten _{F_j}  =
{ \kappa_{B_i}  L_{B_i} A_{B_i} \over 8\pi}  
 +  \frac{   \Lambda\,  (V_{(B_i)} -V_{(F_j)} ) }{ 4\pi (D-2) }
 \end{equation}
Equation (\ref{bubblenc}) can be rearranged to put all the bubble terms with 
 index $B_i$ on one side, and the freely acting terms with index $F_j$ on the other. Since 
 this is true for all choices of $i$ and $j$ and noting an analogous property of
 equation (\ref{spatialnc}) we learn that the combination
  \begin{equation}\label{same}
 L_{j} \ten _{j} - {\Lambda V_{(j)} \over 4\pi (D-2) }   - 
{ L_{j}  \kappa_{j} A_{j} \over 8\pi} =
\mass- {\kappa_H A_H \over 8\pi} -{\Lambda V_{(t)} \over 4\pi (D-2) } \equiv{\cal C}
 \end{equation}
 with the same constant value ${\cal C}$
  for all symmetry directions, and where 
the expression for $L_{F_j} \ten _{F_j}$ is obtained by setting $A_j =0$.
 We recall that in previous work on AdS black holes with spherical 
 topology \cite{Kastor:2009wy,Kubiznak:2012wp,Dolan:2012jh,Dolan:2011xt}
 the combination  $ M -\Lambda V/4\pi (D-3) $ arose as a relevant 
 thermodynamic quantity. For the planar geometry the dimension dependent 
 coefficient changes and  the  corresponding combination is
   $ M -\Lambda V_{(t)}/4\pi (D-2) $. Interestingly, we see that  analogous
 quantities from the  spatial symmetries also play a role in the brane
 thermodynamics, namely the combination
  $L_i\ten _i  -\Lambda V_{(i)} /4\pi (D-2) $. The space-time  quantities
$Q( \partial/\partial x^J ) -\Lambda V_{(J )}/4\pi (D-2)$
altogether form a set of thermodynamic variables, where the 
  $Q( \partial/\partial x^J )$ are the ADM charges associated with space and time 
  translation symmetries. A question for future work is to
  study the properties of these quantities further, for example by computing 
 the corresponding first laws governing their variations.
 
If the spacetime only has a static Killing field then equation (\ref{smarrtime}) is the only Smarr relation.
Using the trace constraint, this can be rewritten as
\begin{equation}\label{timeonly}
\mass +L_{D-2} \ten_{D-2}  =
 {\kappa_H A_H \over 8\pi} +  \frac{   \Lambda V_{(t)} }{ 4\pi (D-2) } 
\end{equation}
This formula would apply, {\it e.g.} to a localized or ``caged"  black hole with spherical horizon topology. For example, suppose that all the periodicities $L_i$ of the spatial directions are equal so that the black hole horizon 
is asymptotically spherically symmetric within each constant $r$ plane. All
the coefficients $c_i$ would then be the same  and all the products $L_i \ten_i$ would be equal. The zero trace
condition then implies that this common value is given by $-M/(D-2)$, so that equation (\ref{timeonly}) becomes
\begin{equation}\label{timethree}
(D-3) \mass =
(D-2) {\kappa_H A_H \over 8\pi} +  \frac{   \Lambda V_{(t)} }{ 4\pi  } ,
\end{equation}
which is the same as the Smarr relation as for a static AdS black hole with spherical horizon topology \cite{Kastor:2009wy}. 

The Smarr relation (\ref{timeonly}) has the puzzling feature that $L_{D-2} \ten_{D-2}$
is singled out, which can be traced back to the gauge choice
$c_{D-2} =0$. It seems to us that there is another piece of physics that needs
to be used to get rid of this apparent gauge dependence. For example, we have
just argued that the additional assumption of spherical symmetry asymptotically on constant $r$ slices
allows one to reach a more satisfying result. It is plausible that specifying the symmetries of
the black hole allows one to more generally determine $L_{D-2} \ten_{D-2}$ in terms
of $\mass$. This remains a question for future work.

 \section{Discussion and Conclusion}\label{end}
In this paper we have studied the implications of spatial translation symmetries
for static AdS black hole and soliton spacetimes. The 
associated ADM charges, which are defined in the context of the Hamiltonian
formulation of general relativity, agree with the corresponding components of the AdS boundary 
stress tensor. We have shown that the sum of the ADM mass and tensions of
a solution is zero.
The set of Smarr formulas obtained provide relations between the
mass, tensions, horizon and bubble areas, as well as a  set of thermodynamic 
volumes.  These relations display a duality between  black holes and bubbles,
in which the mass and black hole area are interchanged with the bubble tension
and bubble area. Given the interpretation as deconfined/confined phases of the
dual gauge theory, it would be of interest to further understand how the black hole/bubble
symmetry manifests in the boundary gauge theory. A related question is to sort out
what the thermodynamic volumes correspond to in the CFT. Since in general the bulk Killing
fields do not commute, it is tempting to conjecture that these ``renormalized" volumes
relate to integrals of (possibly non-commuting) gauge fields around loops in the boundary. 

Another area for further research is to study more general asymptotics, for example,
when the fall-off coefficients depend on the planar coordinates $x^\alpha$. Further,
the AdS soliton-black string solution given in \cite{Haehl:2012tw} 
and the numerical droplet type black hole solutions given in \cite{Figueras:2011va}
do not have the  asymptotically AdS form of solutions considered here, since the black hole
horizon extends to infinity. A question for future consideration is how to define the
ADM charges of these spacetimes and to derive the corresponding Smarr relations.

\vspace{0.2cm}

\section{Acknowledgments}
This work was supported in part by  NSF grant PHY-0555304.
JT thanks the Centro de Ciencias de Benasque Pedro Pascual, where this work was initiated, for their hospitality.


\newpage

\begin{appendices}
\section{Appendix: Killing potential}
\numberwithin{equation}{section}

\noindent We use perturbation theory to solve for the Killing potential in the asymptotic region of the spacetime. Recalling the defintion of the antisymmetric Killing potential, one can write
\begin{equation}
\frac{1}{\sqrt{g}}\partial_a \left(\sqrt{g}\omega^{ab}\right)=K^b ~~~.
\end{equation}

\noindent For definitness, we take the Killing field to be the generator of time translations. The metric functions in (\ref{adsfalloff}) depends solely on $r$, and hence the Killing potential components are consequently assumed to depend only on $r$. We moreover write the asymptotic metric as a perturbative expansion off AdS, and assume a similar expansion to exist for $\omega^{rt}$
\begin{align}
g_{ab}=\bar{g}_{ab}+ \kappa \gamma_{ab} +\mathcal{O}(\kappa^2) \quad , \quad \omega^{rt}(r)=\omega^{rt}_{(0)}+ \kappa \omega^{rt}_{(1)}+\mathcal{O}(\kappa^2) ~~~.
\end{align}

\noindent Expanding the metric determinant then yields
\begin{align}
\sqrt{-g}=\sqrt{-\bar{g}} \left(1+ \kappa \frac{\gamma}{2}+\mathcal{O}(\kappa^2)\right) \quad , \quad \gamma \equiv \bar{g}^{ab}\gamma_{ab} ~~~.
\end{align}
The solution for the zeroth order equation is that for pure AdS, so  $\omega^{rt}_{(1)}$ is the solution to
\begin{align}
\left(\frac{d}{dr}+\frac{D-2}{r}\right)\omega^{rt}_{(1)}=-\frac{r}{2(D-1)}\frac{d\gamma}{dr} ~~~. 
\end{align}
From (\ref{adsfalloff}), we find $\gamma=\frac{c_r+\sum_i c_i-c_t}{2r^{D-1}}$, and finally
\begin{align}
\omega^{rt}_{(1)}=\frac{c_r+\sum_i c_i-c_t}{2r^{D-2}}\ln{r}+\frac{a}{r^{D-2}},
\end{align}

Where $a$ is an integration constant to be determined by appropriate boundary conditions on the Killing potential. Recalling that the leading contribution to the area element in the Komar integral goes as $r^{-(D-2)}$, we see that the first term in the above equation diverges logarithmically at infinity. Since the Komar integral relation is a rewriting of 
  Einstein equation, the sum of the fall-off coefficients has to vanish
  to have solutions with the assumed asymptotics. That is,
  the trace of the metric perturbations vanishes on solutions,
\begin{align}
c_r+\sum_{i=1}^{D-2} c_i-c_t=0
\end{align}

\end{appendices}
 

\end{document}